\documentclass[12pt,a4paper]{jpconf}
\usepackage{graphics,graphicx}
\usepackage{amssymb,amsmath}
\begin{document}
\title{Implications of an additional scale on leptogenesis}

\author{D Aristizabal Sierra$^{a,}$\footnote{Talk given by D. Aristizabal
    Sierra at the Discrete'08 Symposium,  11-16 Dec. 2008,
    Valencia-Spain.}, L A Mu\~noz$^b$ and E Nardi$^{a,b}$}

\address{$^a$ INFN, Laboratori Nazionali di Frascati,C.P. 13, I00044
  Frascati, Italy}

\address{$^b$ Instituto de F\'\i sica, Universidad de Antioquia,
  A.A.{\it{1226}}, Medell\'\i n, Colombia}

\ead{daristi@lnf.infn.it, lmunoz@fisica.udea.edu.co, 
  enrico.nardi@lnf.infn.it}

\begin{abstract}
  We consider variations of the standard leptogenesis picture arising
  from the presence of an additional scale related to the breaking of
  a $U(1)_X$ abelian flavor symmetry. We show that quite generically
  the presence of an additional energy scale might introduce new
  qualitative and quantitative changes on leptogenesis. Especially
  interesting is the possibility of having succesful TeV leptogenesis
  with a vanishing total CP violating asymmetry.  By solving the
  corresponding Boltzmann equations it is shown that these kind of
  scenarios encounters no difficulties in generating the Cosmic baryon
  asymmetry.
\end{abstract}
\section{Introduction}
\label{sec:int}
From observations of light element abundances and of the Cosmic
microwave background radiation \cite{Hinshaw:2008kr} the Cosmic 
baryon asymmetry,
\begin{equation}
  \label{eq:baryon-asymm}
  {\cal Y}_B=(8.75\pm 0.23)\times 10^{-10}\,,
\end{equation}
can be inferred. The conditions for a dynamical generation of this
asymmetry (baryogenesis) are well known \cite{Sakharov:1967dj} and
depending on how they are realized different scenarios for
baryogenesis can be defined (see ref. \cite{Dolgov:1991fr} for a
througout discussion).

Leptogenesis \cite{Fukugita:1986hr} is a scenario in which an initial
lepton asymmetry, generated in the out-of-equilibrium decays of heavy
standard model singlet Majorana neutrinos ($N_\alpha$), is partially
converted in a baryon asymmetry by anomalous sphaleron
interactions~\cite{Kuzmin:1985mm} that are standard model processes.
Singlet Majorana neutrinos are an essential ingredient for the
generation of light neutrino masses through the seesaw mechanism
\cite{Minkowski:1977sc}.  This means
that if the seesaw is the source of neutrino masses then qualitatively
leptogenesis is unavoidable.  Consequently, whether the baryon
asymmetry puzzle can be solved within this framework turn out to be a
quantitative question. This has triggered a great deal of interest on
quantitative analysis of the standard leptogenesis model and indeed a
lot of progress during the last years have been achieved (see
ref. \cite{Davidson:2008bu} for details).

Here we focus on variations of the standard leptogenesis picture which
can arise if, apart from the lepton number breaking scale ($M_N$), an
additional energy scale, related to the breaking of a new symmetry,
exist. We consider a simple realization of this idea in which at an
energy scale of the order of the lepton number violating scale the
tree level coupling linking light and heavy neutrinos is forbiden by
an exact $U(1)_X$ flavor symmetry which below (or above) $M_N$ becomes
spontaneously broken by the vacuum expectation value, $\sigma$, of a
standard model singlet scalar field $S$ and involves heavy vectorlike
fields $F_a$.  As will be discussed, according to the relative size of
the relevant scales of the model ($M_{N_\alpha}$, $\sigma$,
$M_{F_a}$), different scenarios for leptogenesis can be defined
\cite{AristizabalSierra:2007ur}.  Of particular interest is the case
in which the total CP violating (CPV) asymmetries in the decays and
scatterings of the singlet seesaw neutrinos vanish. As we
will discuss further on two remarkable features distinguish this
scenario \cite{amn}: ($a$) Flavor effects are entirely responsible for successful
leptogenesis; ($b$) leptogenesis can be lowered down to the TeV scale.

The rest of this paper is organized as follows: In section
\ref{sec:model} we briefly discuss the model while in
section~\ref{sec:different-possibilities} we discuss two particular
realizations of our scheme paying special attention to the {\it purely
  flavored leptogenesis} (PFL) case for which we analyse the evolution of
the generated lepton asymmetry by solving the corresponding Boltzmann
equations (BE).
\section{The model}
\label{sec:model}
The model we consider here \cite{AristizabalSierra:2007ur} is a simple
extension of the standard model containing a set of $SU(2)_L\times U(1)_Y$
fermionic singlets, namely three right-handed neutrinos ($N_\alpha = N_{\alpha
  R} + N_{\alpha R}^c$) and three heavy vectorlike fields ($F_a=F_{aL} +
F_{aR}$). In addition, we assume that at some high energy scale, taken to be
of the order of the leptogenesis scale $M_{N_1}$, an exact $U(1)_X$ horizontal
symmetry forbids direct couplings of the lepton $\ell_i$ and Higgs $\Phi$
doublets to the heavy Majorana neutrinos $N_\alpha$.  At lower energies,
$U(1)_X$ gets  spontaneously broken by the vacuum expectation value (vev)
$\sigma$ of a $SU(2)$ singlet scalar field $S$.  Accordingly, the Yukawa
interactions of the high energy Lagrangian read
\begin{equation}
  \label{eq:lag}
  -{\cal L}_Y = 
  \frac{1}{2}\bar{N}_{\alpha}M_{N_\alpha}N_{\alpha} +
  \bar{F}_{a}M_{F_a}F_{a} +
  h_{ia}\bar{\ell}_{i}P_{R}F_{a}\Phi + 
\bar{N}_{\alpha}
\left(  \lambda_{\alpha a} +   \lambda^{(5)}_{\alpha a}\gamma_5\right) 
     F_{a}S 
  + \mbox{h.c.}   
\end{equation}
We use Greek indices $\alpha,\beta\dots =1,2,3$ to label the heavy Majorana
neutrinos, Latin indices $a,b\dots =1,2,3$ for the vectorlike messengers, and
$i, j, k, \dots$ for the lepton flavors $e,\mu,\tau$.  Following reference
\cite{AristizabalSierra:2007ur} we chose the simple $U(1)_X$ charge
assignments $X(\ell_{L_i},F_{L_a},F_{R_a})=+1$, $X(S)=-1$ and
$X(N_{\alpha},\Phi)=0$.  This assignment is sufficient to enforce the absence
of $\bar N \ell \Phi$ terms, but clearly it does not constitute an attempt to
reproduce the fermion mass pattern, and accordingly we will also avoid
assigning specific charges to the right-handed leptons and quark fields that
have no relevance for our analysis.  The important point is that it is likely
that any flavor symmetry (of the Froggatt-Nielsen type) will forbid the the
same tree-level couplings, and will reproduce an overall model structure
similar to the one we are assuming here. Therefore we believe that our
results, that are focused on a new realization of the leptogenesis mechanism,
can hint to a general possibility that could well occur also in a complete
model of flavor.

As discussed in~\cite{AristizabalSierra:2007ur}, depending on the
hierarchy between the relevant scales of the model
($M_{N_1},\,M_{F_a},\,\sigma$), quite different {\it scenarios} for
leptogenesis can arise. Here we will concentrate on two cases: ($i$)
The standard leptogenesis case ($M_F,\sigma\gg M_N$); ($ii$) the PFL
case ($\sigma < M_{N_1}< M_{F_a}$) that is, when the flavor symmetry
$U(1)_X$ is still unbroken during the leptogenesis era and at the same
time the messengers $F_a$ are too heavy to be produced in $N_1$ decays
and scatterings, and can be integrated away \cite{amn}.

As is explicitly shown by the last term in eq.~(\ref{eq:lag}), in
general the vectorlike fields can couple to the heavy singlet
neutrinos via scalar and pseudoscalar couplings, In
ref.~\cite{AristizabalSierra:2007ur} it was assumed for simplicity a
strong hierarchy $\lambda\gg \lambda^{(5)}$ so $\lambda^{(5)}$ was
neglected.  However, in all the relevant quantities (scatterings, CP
asymmetries, light neutrino masses) at leading order the scalar and
pseudoscscalar couplings always appear in the combination $\lambda +
\lambda^{(5)}$, and thus such an assumption is not necessary.  The
replacement $\lambda \to \lambda + \lambda^{(5)}$ would suffice to
include in the analysis the effects of both type of interactions.

\subsection{Effective seesaw and light neutrino masses}
\label{sec:nmgen}
\begin{figure}[t]
  \centering
  \includegraphics[width=8cm,height=2.5cm]{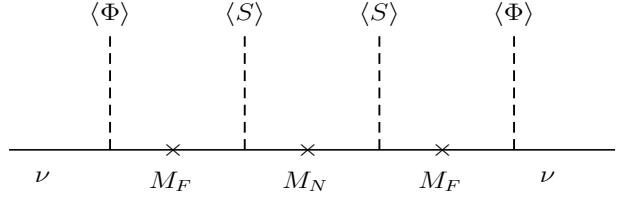}
  \caption{Effective mass operator responsible for neutrino mass generation}
  \label{fig:neutrino-massmatrix}
\end{figure}
After $U(1)_X$ and electroweak symmetry breaking
the set of Yukawa interactions in (\ref{eq:lag}) generate light
neutrino masses through the effective mass operator shown in figure
\ref{fig:neutrino-massmatrix}. The resulting mass matrix can be written as
\cite{AristizabalSierra:2007ur}
\begin{equation}
  \label{eq:nmm}
  -{\cal M}_{ij}=
  \left[
    h^{*}\frac{\sigma}{M_{F}}\lambda^{T}\frac{v^{2}}
    {M_{N}}\lambda\frac{\sigma}{M_{F}}h^{\dagger}
  \right]_{ij}
  = \left[
    \tilde{\lambda}^{T}\frac{v^{2}}{M_{N}}\tilde{\lambda}
  \right]_{ij} \,.
\end{equation}
Here we have introduced the seesaw-like couplings
\begin{equation}
  \label{eq:seesaw-couplings}
  \tilde{\lambda}_{\alpha i} = 
  \left(
    \lambda \frac{\sigma}{M_F}h^\dagger
  \right)_{\alpha i}\,.
\end{equation}
Note that, in contrast to the standard seesaw, the
neutrino mass matrix is of fourth order in the {\it fundamental}
Yukawa couplings ($h$ and $\lambda$) and due to the factor
$\sigma^2/M_F^2$ is even more suppressed.
\section{Different scenarios for leptogenesis}
\label{sec:different-possibilities}
In this section we discuss the features of each one of the
cases we previously mentioned and derive expressions for the CP asymmetries.
Henceforth we will use the following notation for the different mass ratios:
\begin{equation}
  \label{eq:mass-ratios}
  z_{\alpha}=\frac{M_{N_\alpha}^2}{M_{N_1}^2},\qquad \omega_{a}=
  \frac{M_{F_a}^2}{M_{F_1}^2}, \qquad r_a=\frac{M_{N_1}}{M_{F_a}}.
\end{equation}
\subsection{The standard leptogenesis case}
\label{sec:standard}
\begin{figure}[t]
  \centering
  \includegraphics[width=10cm,height=2.5cm]{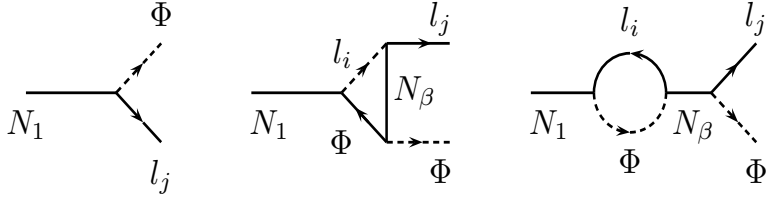}
  \caption{Diagrams responsible for the CP violating asymmetry in the standard case}
  \label{fig:case0}
\end{figure}

When the masses of the heavy fields $F_a$ and the $U(1)_X$ symmetry breaking
scale are both larger than the Majorana neutrino masses ($M_F\,, \sigma >
M_N$) there are no major differences from the standard Fukugita-Yanagida
leptogenesis model \cite{Fukugita:1986hr}.  After integrating out the $F$ fields one
obtains the standard seesaw Lagrangian containing the effective operators
$\tilde\lambda_{\alpha i} \bar N_\alpha l_i \Phi$ with the seesaw couplings
$\tilde\lambda_{\alpha i}$ given in eq.~(\ref{eq:seesaw-couplings}).  The right
handed neutrino $N_1$ decays predominantly via 2-body channels as shown in
fig.~\ref{fig:case0}.  This yields the standard results that for convenience
we recall here.  The total decay width is $\Gamma_{N_1}= \left({M_{N_1}}/{16
    \pi}\right) (\tilde\lambda\tilde\lambda^\dagger)_{11}$ and the sum of the
vertex and self-energy contributions to the $CP$-asymmetry for $N_1$ decays
into the flavor $l_j$ reads \cite{Covi:1996wh}
\begin{equation}
  \label{eq:6}
  \epsilon_{N_1\to l_j}=\frac{1}{8\pi (\tilde\lambda \tilde\lambda^\dagger)_{11}}
  \sum_{\beta\neq 1}{\mathbb I}\mbox{m}
  \left\{\tilde\lambda_{\beta j}\tilde\lambda^*_{1j} \left[
      (\tilde\lambda\tilde\lambda^\dagger)_{\beta1}    \tilde F_1(z_\beta) +
      (\tilde\lambda\tilde\lambda^\dagger
      )_{1\beta}
      \tilde F_2(z_\beta)
    \right]\right\}\,,
\end{equation}
where 
%
\begin{equation}
  \label{eq:tildeF}
      \tilde F_1(z )= \frac{\sqrt{z }}{1-z } +
  \sqrt{z }\left(1-(1+z )\ln
    \frac{1+z }{z }\right),   
\qquad     \tilde F_2(z)=\frac{1}{1-z}. 
\end{equation}
%
At leading order in $1/z_\beta$ and after summing over all leptons $l_j$,    
eq.~(\ref{eq:6}) yields for the total asymmetry: 
\begin{equation}
        \label{eq:8} 
\epsilon_{N_1}=\frac{3}{16\pi (\tilde\lambda \tilde\lambda^\dagger)_{11}}
\sum_{\beta}
{\mathbb I}\mbox{m}
\left\{\frac{1}{\sqrt{z_\beta}}
       (\tilde\lambda\tilde\lambda^\dagger)_{\beta1}^2 \right\}. 
\end{equation}
where the sum over the heavy neutrinos has been extended to include also $N_1$
since  for $\beta=1$  the corresponding combination of couplings is real.

In the hierarchical case $M_{N_1}\ll M_{N_{2,3}}$  
the size of the total asymmetry in (\ref{eq:8}) is bounded by 
the Davidson-Ibarra limit  \cite{Davidson:2002qv}
\begin{equation}
\label{eq:DI}
|\epsilon_{N_1}|\leq \frac{3}{16\pi}\frac{M_{N_1}}{v^2} \,(m_{\nu_3}-m_{\nu_1})
\lesssim \frac{3}{16\pi}\frac{M_{N_1}}{v^2} \frac{\Delta m^2_{\text{atm}}}{2m_{\nu_3}}\,, 
\end{equation}
where $m_{\nu_i}$ (with $m_{\nu_1} < m_{\nu_2} < m_{\nu_3}$) are the
light neutrinos mass eigenstates and $\Delta m^2_{\text{atm}}\sim
2.5\times 10^{-3}\,$eV$^2$ is the atmospheric neutrino mass difference
\cite{Schwetz:2008er}.  It is now easy to see that (\ref{eq:DI})
implies a lower limit on $M_{N_1}$.  The amount of $B$ asymmetry that
can be generated from $N_1$ dynamics can be written as:
\begin{equation}
\label{eq:etaB}
\frac{n_B}{s}=-\kappa_s\,\epsilon_{N_1}\,\eta  ,
\end{equation} 
where $\kappa_s\approx 1.3\times 10^{-3}$ accounts for the dilution of
the asymmetry due to the increase of the Universe entropy from the
time the asymmetry is generated with respect to the present time,
$\eta $ (that can range between 0 and 1, with typical values
$10^{-1}-10^{-2}$) is the {\it efficiency factor} that accounts for
the amount of $L$ asymmetry that can survive the washout process.
Assuming that $\epsilon_{N_1}$ is the main source of the $B-L$
asymmetry~\cite{Engelhard:2006yg}, eqs.~(\ref{eq:DI}) and (\ref{eq:etaB}) in
addition from the oserved baryon asymmetry eq. (\ref{eq:baryon-asymm})
yield:
\begin{equation}
  \label{eq:M1limit0}
  M_{N_1} \gtrsim 10^{9}\, 
\frac{m_{\nu_3}}{\eta\, \sqrt{\Delta m^2_{\text{atm}}}} \, {\rm GeV}.
\end{equation}
This limit can be somewhat relaxed depending on the specific initial
conditions~\cite{Giudice:2003jh} or when flavor effects are
included~\cite{Nardi:2006fx,Abada:2006fw,Abada:2006ea,JosseMichaux:2007zj} but the main point remains,
and that is that the value of $M_{N_1}$ should be well above the electroweak
scale.
\subsection{Purely flavored leptogenesis case}
\label{sec:PFL}
\begin{figure}[t]
\begin{center}
\includegraphics[width=10cm,height=3cm]{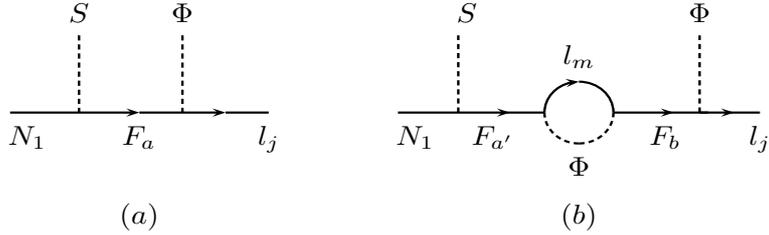}
\end{center}
\caption{Feynman diagrams responsible for the CPV asymmetry.}
\label{fig:fig0}
\end{figure}

Differently from standard leptogenesis in the present case, since $M_F >
M_{N_1}$, two-body $N_1$ decays are kinematically forbidden.  However, via
off-shell exchange of the heavy $F_a$ fields, $N_1$ can decay to the three
body final states $S\Phi l$ and $\bar S\bar\Phi \bar l$. The corresponding
Feynman diagram is depicted in figure \ref{fig:fig0}$(a)$.  At leading order
in $r_a=M_{N_1}/M_{F_a}$, the total decay width reads
\cite{AristizabalSierra:2007ur}
\begin{equation}
  \label{eq:total-decay-width}
 \Gamma_{N_1}\equiv  
\sum_j \Gamma(N_1\to S\Phi l_j + \bar S\bar\Phi \bar l_j) =\frac{M_{N_1}}{192\pi^3}
  \left(
    \frac{M_{N_1}}{\sigma}  
  \right)^2
  (\tilde{\lambda}\tilde{\lambda}^\dagger)_{11}\,.
\end{equation}

As usual, CPV asymmetries in $N_1$ decays arise from the interference between
tree-level and one-loop amplitudes.  As was noted
in~\cite{AristizabalSierra:2007ur}, in this model at one-loop there are no
contributions from vertex corrections, and the only contribution to the CPV
asymmetries comes from the self-energy diagram~\ref{fig:fig0}$(b)$.  Summing
over the leptons and vectorlike fields running in the loop, at leading order in
$r_a$ the CPV asymmetry for $N_1$ decays into leptons of flavor $j$ can be
written as
\begin{equation}
  \label{eq:cp-violating-asymm}
 \epsilon_{1j}  \equiv    \epsilon_{N_1\to\ell_j} = 
  \frac{3}{128\pi}
  \frac{\sum_{m} \mathbb{I}\mbox{m} 
    \left[
      \left(
        h r^{2} h^{\dagger}
      \right)_{mj}\tilde{\lambda}_{1m}\tilde{\lambda}^{*}_{1j} 
    \right]}{\left(\tilde{\lambda}\tilde{\lambda}^{\dagger}\right)_{11}}\,.
\end{equation}
Note that since the loop correction does not violate lepton number, the total
CPV asymmetry that is obtained by summing over the flavor of the final state
leptons vanishes~\cite{Kolb:1979qa}, that is $\epsilon_{1}\equiv \sum_j
\epsilon_{1j}=0$.  This is the condition that defines PFL; namely there is no
CPV {\it and} lepton number violating asymmetry, and the CPV lepton flavor
asymmetries are the only seed of the Cosmological lepton and baryon
asymmetries.

It is important to note that the effective couplings $\tilde\lambda$ defined
in eq.~(\ref{eq:seesaw-couplings}) are invariant under the reparameterization
\begin{equation}
  \label{eq:couplingtrans}
  \lambda\to \lambda\cdot (rU)^{-1},\quad
  h^\dagger \to (U r)\cdot h^\dagger\,,
\end{equation}
where $U$ is an arbitrary $3\times 3$ non-singular matrix.  Clearly
the light neutrino mass matrix is invariant under this transformation.
Moreover, also the flavor dependent washout processes, that correspond
to tree level amplitudes that are determined, to a good approximation,
by the effective $\tilde \lambda$ couplings, are left essentially
unchanged.\footnote{The approximation is exact in the limit of
  pointlike $F$-propagators $(s-M^2_F+iM_F\Gamma_F)\to M^2_F$.} On the
contrary, the flavor CPV asymmetries
eq.~(\ref{eq:cp-violating-asymm}), that are determined by loop
amplitudes containing an additional factor of $h r^2 h^\dagger$, get
rescaled as $h r^2 h^\dagger\to h (rUr)^\dagger (rUr) h^\dagger $.
Clearly, this rescaling affects in the same way all the lepton flavors
(as it should be to guarantee that the PFL conditions
$\epsilon_\alpha\equiv \sum\epsilon_{\alpha j}=0$ are not spoiled),
and thus for simplicity we will consider only rescaling by a global
scalar factor $r.U=U.r=\kappa\,I$ (with $I$ the $3\times 3$ identity
matrix) that, for our purposes, is completely equivalent to the more
general transformation~(\ref{eq:couplingtrans}). Thus, while rescaling
the Yukawa couplings through
\begin{equation}
  \label{eq:coupling-rescaling-gen}
  \lambda\to \lambda \,\kappa^{-1},\quad 
  h^\dagger\to\kappa \,h^\dagger\,,
\end{equation}
does not affect neither low energy neutrino physics nor the washout processes,
the CPV asymmetries get rescaled as:
\begin{equation}
  \label{eq:rescaled-CPV-asymm}
  \epsilon_{1j}\to\kappa^2\epsilon_{1j}\,.
\end{equation}
By choosing $\kappa>1$, all the CPV asymmetries get enhanced as
$\kappa^2$ and, being the Cosmological asymmetries generated through
leptogenesis linear in the CPV asymmetries, the final result gets
enhanced by the same factor.  Therefore, for any given set of
couplings, one can always find an appropriate rescaling such that the
correct amount of Cosmological lepton asymmetry is generated.  In
practice, the rescaling factors $\kappa$ cannot be arbitrarily large:
first, they should respect the condition that all the fundamental
Yukawa couplings remain in the perturbative regime; second the size of
the $h$ couplings (and thus also of the rescaling parameter $\kappa$)
is also constrained by experimental limits on lepton flavor violating
decays.

\begin{figure}[t]
\begin{center}
\includegraphics[width=13.0cm,height=2.3cm,angle=0]{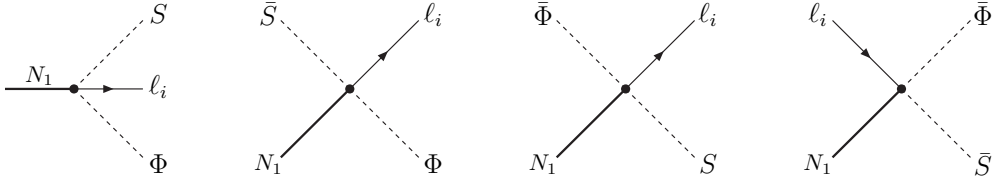}
\end{center}
\caption{Feynman diagrams for $1\leftrightarrow 3$ and
  $2\leftrightarrow 2$ $s$, $t$ and $u$ channel processes 
  after integrating out the heavy vectorlike fields $F_a$.}
 \label{fig:fig1}
 \end{figure}
\subsubsection{Boltzmann Equations}
\label{sec:BE}
In this section we compute the lepton asymmetry by solving the
appropriate BE. In general, to consistently derive the evolution
equation of the lepton asymmetry all the possible processes at a given
order in the couplings have to be included.  In the present case
$1\leftrightarrow 3$ decays and inverse decays, and $2\leftrightarrow
2$ $s$, $t$ and $u$ channel scatterings all occur at the same order in
the couplings and must be included altogether in the BE. The Feynman
diagrams for these processes are shown in Figure~\ref{fig:fig1}.  In
addition, the CPV asymmetries of some higher order multiparticle
reactions involving the exchange of one off-shell $N_1$, also
contribute to the source term of the asymmetries at the same order in
the couplings than the CPV asymmetries of decays and $2\leftrightarrow
2$ scatterings.  More precisely, for a proper derivation of the BE it
is essential that the CPV asymmetries of the off-shell
$3\leftrightarrow 3$ and $2\leftrightarrow 4$ scattering processes are
also taken into account \cite{amn}.

As regards the equation for the evolution of the heavy neutrino density
$Y_{N_1}$,  only the diagrams in fig.~\ref{fig:fig1}, that are of leading order 
in the couplings, are important \cite{amn}
\begin{align}
  \label{eq:BEforN-section}
  \dot{Y}_{N_1} &= 
  -\left(y_{N_1} - 1 \right) \gamma_{\text{tot}}\,, \\
  \label{eq:BEforLA-section}
  \dot{Y}_{\Delta \ell_{i}} &=
  \left(y_{N_1} - 1 \right)\epsilon_{i}\gamma_{\text{tot}} - \Delta y_{i}
  \left[
    \gamma_{i} + \left(y_{N_1} - 1 \right)\gamma^{N_{1}\bar\ell_{i}}_{S\Phi}
  \right]\,, 
\end{align}
Here we have normalized particle densities to their equilibrium
densities $y_a\equiv Y_a/Y_a^{\text{eq}}$ where $Y_a=n_a/s$ with $n_a$
the particle number density and $s$ the entropy density. The time
derivative is defined as $\dot Y = sHz\,dY/dz$ with $z=M_{N_1}/T$ and
$H$ is the Hubble parameter. In the last term of the second equation
we have used the compact notation for the reaction densities
$\gamma^{N_{1}\bar\ell_{i}}_{S\Phi} =\gamma(N_{1}\bar\ell_{i}\to
{S\Phi})$ and in addition we have defined
\begin{eqnarray}
  \label{eq:rates}
    \gamma_{i} &=& 
 \gamma^{N_{1}}_{S\ell_{i}\Phi}+
  \gamma^{N_{1}\bar{S}}_{\Phi\ell_{i}} +
  \gamma^{N_{1}\bar{\Phi}}_{S\ell_{i}} + \gamma_{S\Phi}^{N_{1}\bar\ell_{i}}\,, \\
\gamma_{\text{tot}} &=& \sum_{i=e,\mu,\tau} 
    \gamma_{i}+\bar\gamma_i\,, 
\end{eqnarray}
where in the second equation $\bar\gamma_i$ represents the 
sum of the  CP conjugates of the processes summed in $\gamma_{i}$.
 
Since in this model $N_1$ decays are of the same order in the couplings
than scatterings (that is ${\cal O}(\tilde\lambda^2)$), the appropriate condition
that defines the {\it strong washout} regime in the case at hand reads:
\begin{equation}
  \label{eq:strong-washout}
  \left .\frac{\gamma_{\text{tot}}}{z\,H\,s}\right|_{z\sim 1}>1 \,\qquad {\rm
  (strong\ washout)}, 
\end{equation}
and conversely $\gamma_{\text{tot}}/(z\,H\,s)|_{z\sim 1}<1$ defines the {\it
  weak washout} regime.  Note that this is different from standard
leptogenesis, where at $z\sim 1$ two body decays generally dominate over
scatterings, and  e.g. the condition for the strong washout regime can be 
approximated as $\gamma_{\text{tot}}/(z\,H\,s)|_{z\sim 1}\sim
\Gamma_{N_1}/H|_{z\sim 1}>1$.

\subsubsection{Results}
\label{sec:results}
In this section we discuss a typical example of successful
leptogenesis at the scale of a few TeV.  The example presented is a
general one. No particular choice of the parameters has been
performed, except for the fact that the low energy neutrino data are
reproduced within errors, and that the choice yields an interesting
washout dynamics well suited to illustrate how PFL works.  The
numerical value of the final lepton asymmetry ($Y_{\Delta L} \sim
-7.2\times 10^{-10}$) is about a factor of 3 {\it larger} than what is
indicated by measurements of the Cosmic baryon asymmetry. This is
however irrelevant since, as was already discussed, it would be
sufficient a minor rescaling of the couplings (or a slight change in
the CPV phases) to obtain the precise experimental result.  In the
numerical analysis we have neglected the dynamics of the heavier
singlet neutrinos since the $N_\alpha$ masses are sufficiently hierarchical
to ensure that $N_{2,3}$ related washouts do not interfere with $N_1$
dynamics. Moreover, in the (strong washout) fully flavored regime
(that is effective as long as $T < 10^{9}\,$GeV) the $N_{2,3}$ CPV
asymmetries do not contribute to the final lepton number
asymmetry~\cite{Engelhard:2006yg}.

\begin{figure}[t]
\begin{center}
 {\includegraphics[width=6.5cm,height=5.5cm,angle=0]{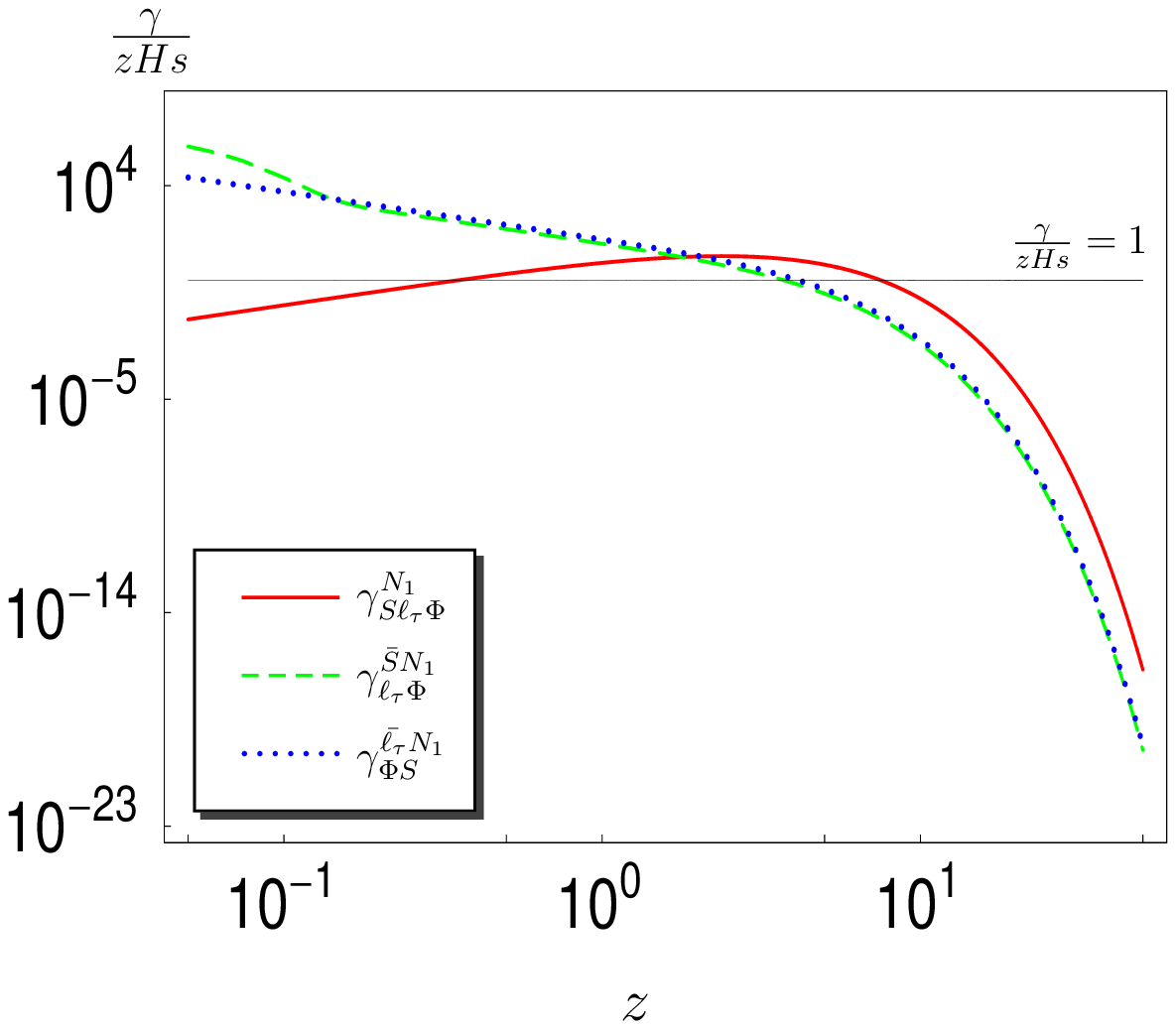}}
\hspace{5mm}
 {\includegraphics[width=6.5cm,height=5.5cm,angle=0]{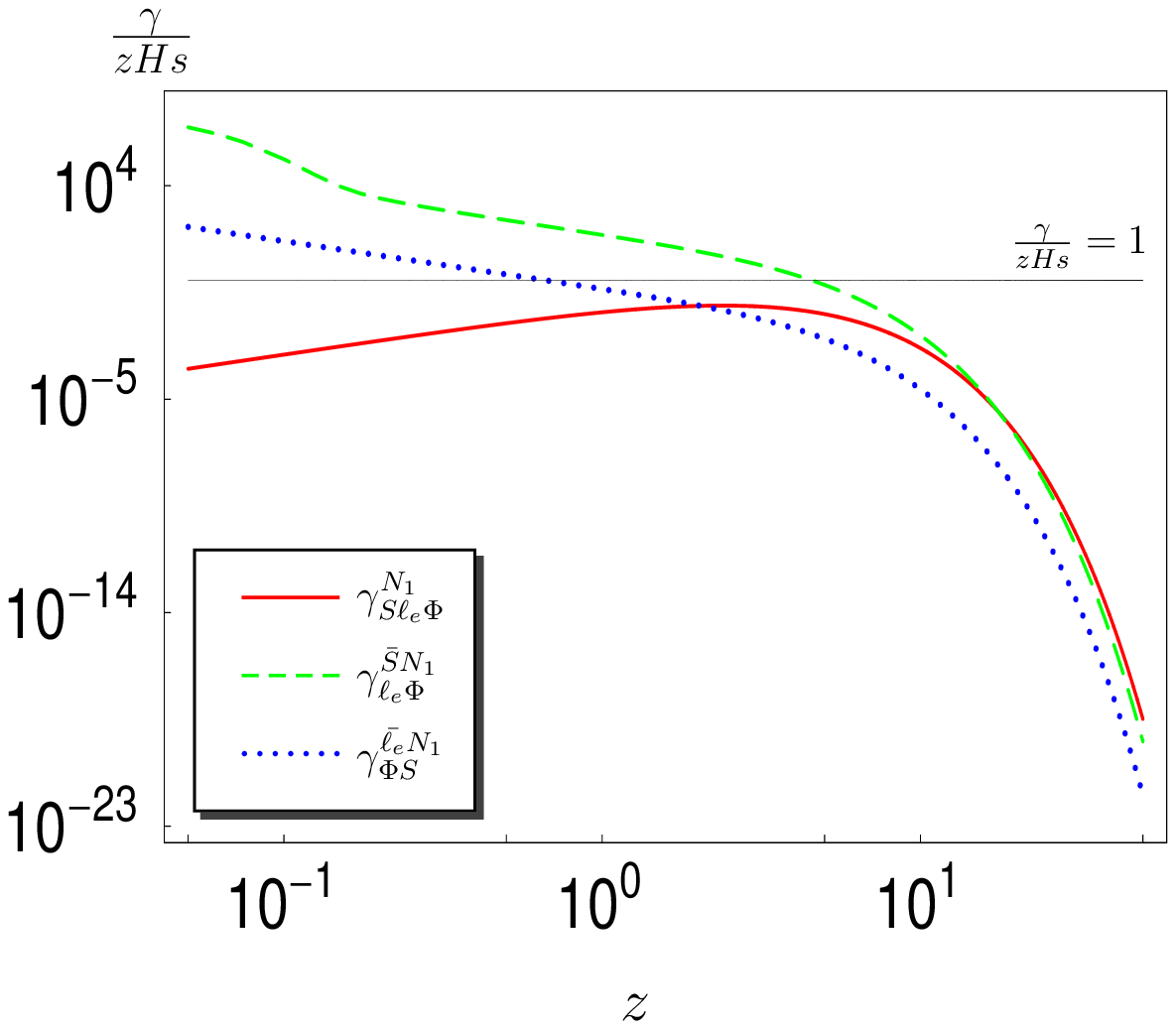}}
\end{center}
\caption{
Reaction densities  normalized to $zHs$  for  $N_1\to S\ell\Phi$ decays
(red solid lines),  $s$-channel $\bar S N_1 
\leftrightarrow \ell \Phi$ scatterings (green dashed lines), and 
$t,u$-channel scatterings in the point-like approximation (blue dotted lines).
Left panel:  $\tau$ flavor. Right panel:  electron flavor. 
}
\label{fig:fig4}
\end{figure}

\begin{figure}[t]
\begin{center}
{\includegraphics[width=6.5cm,height=5.5cm,angle=0]{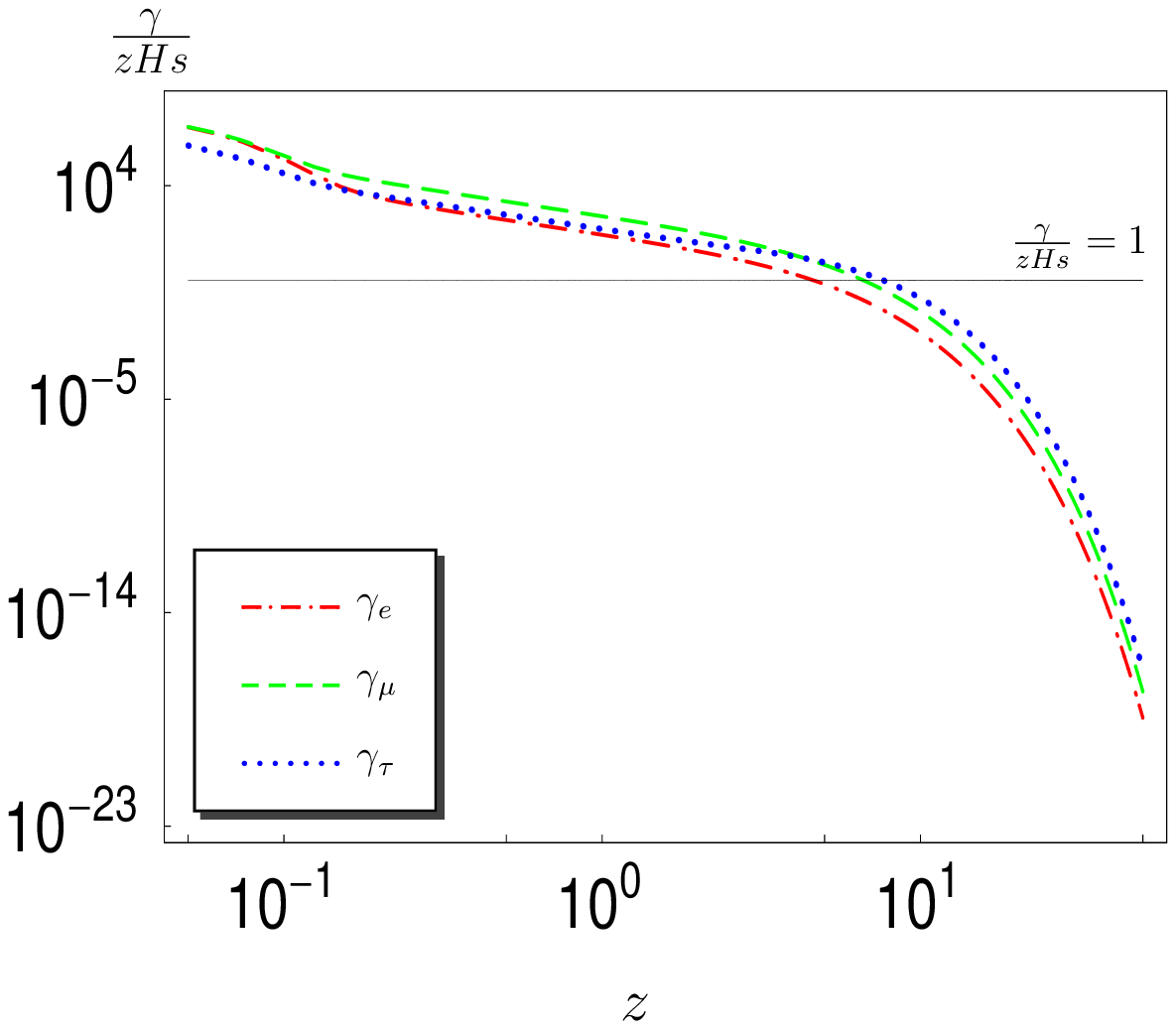}}
\hspace{5mm}
{\includegraphics[width=6.5cm,height=5.5cm,angle=0]{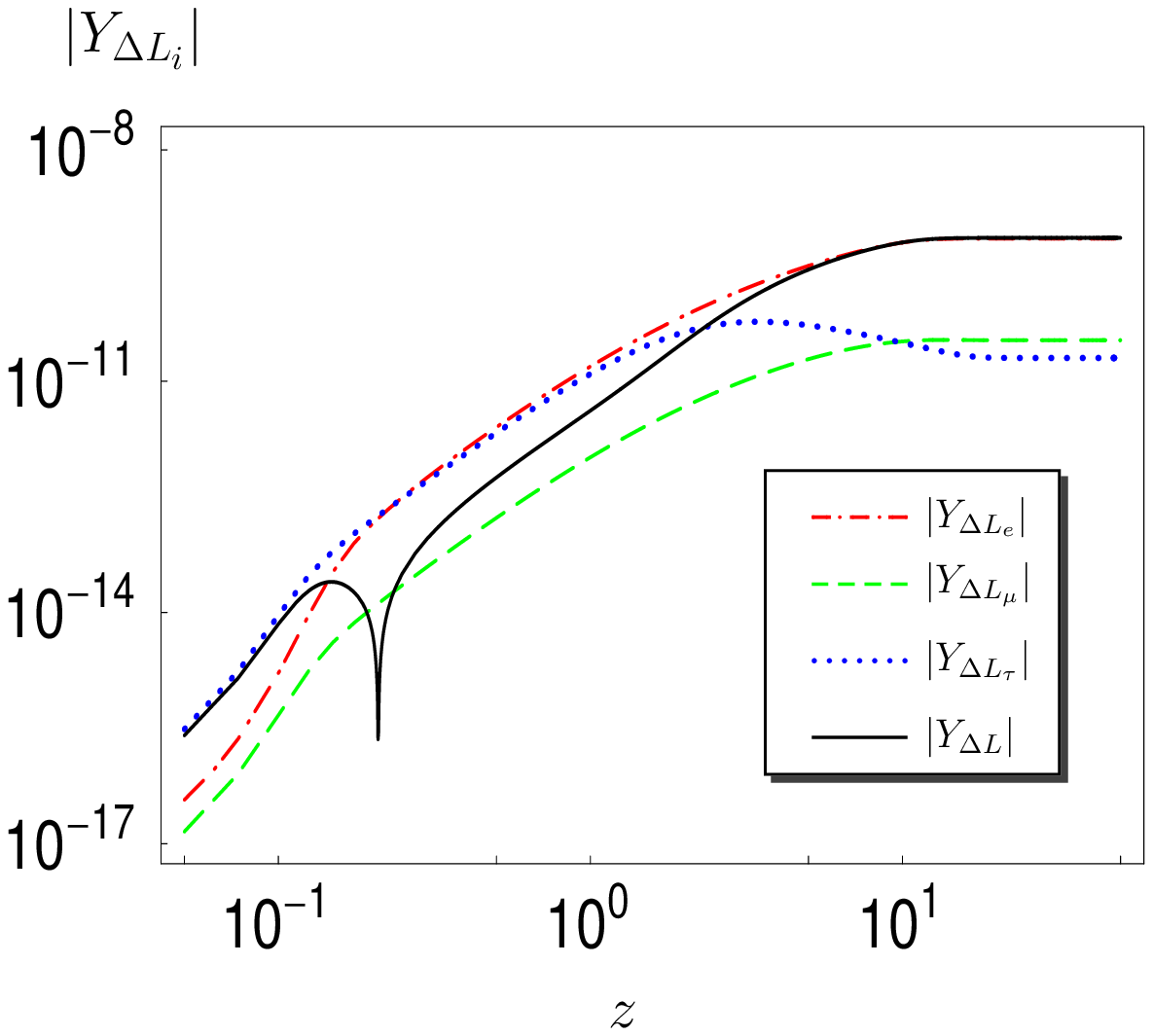}}
\end{center}
\caption{Left panel: the total washout rates for each lepton flavor 
  normalized to $zHs$ as a function of $z$.  Right panel: the
  evolution of the absolute value of the flavored density asymmetries and of
  the lepton number asymmetry (black solid line).  The flavor CPV asymmetries
  are $\epsilon_{1e} = -4.7\times 10^{-4}$, $\epsilon_{1\mu} = -1.9\times
  10^{-4}$ and $\epsilon_{1\tau} = 6.6\times 10^{-4}$.  The final values of
  the asymmetry densities (at $z\gg 1$) are $Y_{\Delta_{\ell_e}} =-7.1\times
  10^{-10}$, $Y_{\Delta_{\ell_\mu}} =-0.3\times 10^{-10}$,
  $Y_{\Delta_{\ell_\tau}} =0.2\times 10^{-10}$.  
}
\label{fig:fig5}
\end{figure}

In figure~\ref{fig:fig4} we show the behavior of the various reaction
densities for decays and scatterings, normalized to $sHz$, as a function of
$z$.  The results correspond to a mass of the lightest singlet
neutrino fixed to $M_{N_1}=2.5\,\text{TeV}$, the heavier neutrino masses are $
M_{N_2}=10\,$TeV and $M_{N_3}=15\,$TeV, and the relevant mass ratios
$r_a=M_{N_1}/M_{F_a}$ for the messenger fields are $r_{1,2,3}=0.1,0.01,0.001$
(the effects of the lightest $F$ resonances can be seen in the $s$-channel
rates in both panels in fig.~\ref{fig:fig4}).  The fundamental Yukawa
couplings $h$ and $\lambda$ are chosen to satisfy the requirement that the
seesaw formula eq.~(\ref{eq:nmm}) reproduces within $2\,\sigma$ the low energy
data on the neutrino mass squared differences and mixing
angles~\cite{Schwetz:2008er}.  Typically, when this requirement is fulfilled,
one also ends up with a dynamics for all the lepton flavors in the strong
washout regime. This is shown in the left panel in fig.~\ref{fig:fig5}
where we present the total rates for the three flavors.

The left panel in fig.~\ref{fig:fig4} refers to the decay and scattering rates
involving the $\tau$-flavor that, in our example, is the flavor more strongly
coupled to $N_1$, and that thus suffers the strongest washout. It is worth
noticing that, due to the fact that in this model scatterings are not
suppressed by additional coupling constants with respect to the decays, the
decay rate starts dominating the washouts only at $z\gtrsim 1$.  The right panel
in fig.~\ref{fig:fig4} depicts the reaction rates for the electron flavor,
that is the more weakly coupled, and for which the strong washout condition
eq.~(\ref{eq:strong-washout}) is essentially ensured by sizeable $s$-channel
scatterings.  Scatterings and decay rates for the $\mu$-flavor are not shown,
but they are in between the ones of the previous two flavors.

The total reaction densities that determine the washout rates for the
different flavors are shown in the first panel in figure~\ref{fig:fig5}.  The
evolution of these rates with $z$ should be confronted with the evolution of
(the absolute value of) the asymmetry densities for each flavor, depicted in
the second panel on the right. Since, as already stressed several times, PFL
is defined by the condition that the sum of the flavor CPV asymmetry vanishes
($\sum_j \epsilon_{1j}=0$), it is the hierarchy between these washout rates
that in the end is the responsible for generating a net lepton number
asymmetry.  In the case at hand, the absolute values of the flavor CPV
asymmetries satisfy the condition $|\epsilon_\mu| < |\epsilon_e| <
|\epsilon_\tau|$, as can be inferred directly by the fact that at $z< 0.1$,
when the effects of the washouts are still negligible, the asymmetry densities
satisfy this hierarchy.  Moreover, since $\epsilon_{\mu,e} < 0$ while
$\epsilon_\tau>0$, initially the total lepton number asymmetry, that is
dominated by $Y_{\Delta_{\ell_\tau}}$, is positive.  As washout effects become
important, the $\tau$-related reactions (blue dotted line in the left panel)
start erasing $Y_{\Delta_{\ell_\tau}}$ more efficiently than what happens for
the other two flavors, and thus the initial positive asymmetry is driven
towards zero, and eventually changes sign around $z=0.2$. This change of sign
corresponds to the steep valley in the absolute value $|Y_{\Delta L}|$ that is
drawn in the figure with a black solid line.  
Note that when all flavors are in  the strong washout regime, 
as in the present case, the condition for the occurrence of 
this `sign inversions'  is simply given by  
$ {\rm max_{j\in e,\mu}}
\left(|\epsilon_j|/|\tilde\lambda_{1j}|^2\right)
\gtrsim \epsilon_\tau/|\tilde\lambda_{1\tau}|^2$. 
From this point onwards, the
asymmetry remains negative, and since the electron flavor is the one that
suffers the weakest washout, $Y_{\Delta_{\ell_e}}$ ends up dominating all the
other density asymmetries.  In fact, as can be seen from the 
right panel in fig.~\ref{fig:fig5}, it is  $Y_{\Delta_{\ell_e}}$
that determines to a large extent the final value of the lepton asymmetry 
$Y_{\Delta  L}=-7.2\times 10^{-10}$.

A few comments are in order regarding the role played by the $F_a$ fields.
Even if $M_{N_1}\ll M_{F_a}$, at large temperatures $z\gg 1$ the tail of the
thermal distributions of the $N_1,\,S$ and $\Phi$ particles allows the
on-shell production of the lightest $F$ states. A possible asymmetry generated
in the decays of the $F$ fields can be ignored for two reasons: first because
due to the rather large $h$ and $\lambda$ couplings $F$ decays occur to a good
approximation in thermal equilibrium, ensuring that no sizeable asymmetry can
be generated, and second because the strong washout dynamics that
characterizes $N_1$ leptogenesis at lower temperatures is in any case
insensitive to changes in the initial conditions.

In conclusion, it is clear from the results of this section that the model
encounters no difficulties to allow for the possibility of generating the
Cosmic baryon asymmetry at a scale of a few TeVs. Moreover, our analysis
provides a concrete example of PFL, and shows that the condition
$\epsilon_1\neq 0$ is by no means required for successful leptogenesis.
\section{Conclusions}
Variations of the standard leptogenesis picture can arise from the
presence of an additional energy scale different from that of lepton
number violation. Quite generically the resulting scenarios are
expected to yield qualitative and quantitative changes on
leptogenesis. Here we have considered what we regard as the simplest
possibility namely, the presence of an abelian flavor symmetry
$U(1)_X$. We have described two possible scenarios within this framework
and have explored their implications for leptogenesis.

We have found that as long as the abelian flavor symmetry energy
scales remain above the lepton number violating scale neither
qualitative nor quantitative differences with the standard
leptogenesis model arise. Conversely if the $U(1)_X$ is unbroken
during the leptogenesis era and the messengers fields $F_a$ are too
heavy to be produced on-shell in $N_1$ decays {\it purely flavored
  leptogenesis} at the TeV scale results. By solving the corresponding
BE we have shown that within this scenario the non-vanishing of the
CPV lepton flavor asymmetries in addition to the lepton and flavor
violating washout processes occuring in the plasma provide the
necessary ingredients to generate the Cosmic baryon asymmetry.
Accordingly, if below the leptogenesis scale new energy scales are
present -as might be expected- the interplay between these scales
could have a quite interesting impact on leptogenesis.

\end{document}